\documentclass{article}

\usepackage{arxiv}

\usepackage[utf8]{inputenc} 
\usepackage[T1]{fontenc}    
\usepackage{hyperref}       
\usepackage{url}            
\usepackage{booktabs}       
\usepackage{amsfonts}       
\usepackage{nicefrac}       
\usepackage{microtype}      
\usepackage{lipsum}
\usepackage{natbib}

\usepackage{amsmath,amssymb}
\usepackage{changepage}
\usepackage{graphicx}

\title{Bet-hedging strategies in expanding populations}

\author{
  Paula Villa Mart{\'\i}n \\
  Biological Complexity Unit,\\
  Okinawa Institute of Science and Technology Graduate University, \\
  Onna, Okinawa 904-0495, Japan \\
   \And
 Miguel A. Mu\~noz \\
  Departamento de Electromagnetismo y F{\'i}sica de la Materia \\
  and Instituto Carlos I de F{\'i}sica Te{\'o}rica y
  Computacional, \\
  Universidad de Granada, Granada, Spain \\
  \AND
  Simone Pigolotti \\
  Biological Complexity Unit,\\
  Okinawa Institute of Science and Technology Graduate University, \\
  Onna, Okinawa 904-0495, Japan \\
  \texttt{simone.pigolotti$@$oist.jp} \\
}

\begin{document}
\maketitle

\begin{abstract}
In ecology, species can mitigate their extinction risks in uncertain
  environments by diversifying individual phenotypes.  This
  observation is quantified by the theory of bet-hedging, which
  provides a reason for the degree of phenotypic diversity observed
  even in clonal populations. The theory of bet-hedging in well-mixed
  populations is rather well developed. However, many species
  underwent range expansions during their evolutionary history, and
  the importance of phenotypic diversity in such scenarios still needs
  to be understood.  In this paper, we develop a theory of bet-hedging
  for populations colonizing new, unknown environments that fluctuate
  either in space or time. In this case, we find that bet-hedging is
  a more favorable strategy than in well-mixed populations.
  For slow rates of variation, temporal and
  spatial fluctuations lead to different outcomes. In spatially
  fluctuating environments, bet-hedging is favored compared to temporally
  fluctuating environments. In the limit of frequent environmental
  variation, no opportunity for bet-hedging exists, regardless of the
  nature of the environmental fluctuations. For the same
model, bet-hedging is never an advantageous strategy in the well-mixed
case, supporting the view that range expansions strongly promote
diversification. These conclusions are
  robust against stochasticity induced by finite population sizes. Our
  findings shed light on the importance of phenotypic heterogeneity in
  range expansions, paving the way to novel approaches to understand
  how biodiversity emerges and is maintained.
\end{abstract}

\keywords{Population dynamics \and Environmental variability \and Range expansion \and Fisher waves}

\newpage
\section{Introduction}
The dynamics and evolutionary history of many biological
species, from bacteria to humans, are characterized by invasions and
expansions into new territory.  The effectiveness of such expansions
is crucial in determining the ecological range and therefore the
success of a species.  A large body of
observational \citep{ramachandran2005support,duckworth2008adaptive} and
experimental
\citep{wolfe1989migration,hallatschek2007genetic,mayor2016front,fu2018spatial}
literature indicates that evolution and selection of species
undergoing range expansions can be dramatically different from that of
other species resident in a fixed habitat.  Theoretical studies of
range expansions based on
the Fisher-Kolmogorov equation \citep{Fisher,Kolmogorov} or variants
\citep{neubert2000demography,barton2012risky} also support this idea.
Adaptive dispersal
strategies \citep{duckworth2008adaptive} and small population sizes
at the edges of expanding fronts
\citep{waters2013founder,hallatschek2008gene} are among the main
reasons for such differences.

Range expansions often occur in non-homogeneous and fluctuating
environments.  Under such conditions, it is possible to mathematically
predict the expansion velocity of a community of phenotypically
identical individuals
\citep{shigesada1979spatial,shigesada1986traveling,
  shigesada1997biological,hastings2005spatial,schreiber2009invasion,dewhirst2009dispersal}.
However, diversity among individuals is expected to play an important
positive role when populations expand in fluctuating environments.
For instance, diverse behavioral strategies help animal populations to
overcome different invasion stages and conditions
\citep{wolf2012animal,sih2012ecological,chapple2012can,carere2013animal}.
Analyses of phenotypic diversity in motile cells suggest that it also
may lead to a selective advantage at a population level
\citep{frankel2014adaptability,dufour2014limits,dufour2016direct}.
Although several studies have tackled the problem of how individual
variability affects population expansion
\citep{neubert2000demography,barton2012risky,fogarty2011social,fu2018spatial,
  ben2000cooperative,keller1971traveling,ben2000cooperative,
  lin2014development,emako2016traveling}, systematic and predictive
theory is still lacking \citep{carere2013animal}.

Phenotypic diversification is often interpreted as a bet-hedging
strategy, spreading the risk of uncertain environmental conditions
across different phenotypes adapted to different environments
\citep{veening2008bistability,kussell2005phenotypic,
  wolf2005diversity,wolf2005microbial,solopova2014bet,
  stumpf2002herpes,rouzine2015evolutionary,
  childs2010evolutionary,hidalgo2016environmental,hopper1999risk}.
Since its formalization in the context of information theory and
portfolio diversification \citep{kelly2011new,fernholz1982stochastic},
a large number of works have explored the applicability of bet-hedging
in evolutionary game theory
\citep{smith1988evolution,nowak2006evolutionary,harmer1999game,parrondo2000new}
and ecology
\citep{de2011bet,williams2011paradoxical,comins1980evolutionarily,hamilton1977dispersal,jansen1998populations}.
Few studies have explored the benefits of bet-hedging in
spatially structured ecosystems
\citep{hidalgo2015stochasticity,rajon2009spatially}.

In this paper, we study how bet-hedging strategies can aid populations
in invading new territories characterized by fluctuating
environments. In particular, we analyze the effect of spatial
expansion, different types of environmental heterogeneity, and
demographic stochasticity on development of bet-hedging strategies for
a population front evolving according to a Fisher wave. 

By employing mathematical as well as extensive computational analyses,
we find that the advantage of bet-hedging in range expansions depends
on the rate of environmental variation. In particular, bet-hedging is
more convenient for infrequently varying environments, whereas its
advantages vanish for frequent environmental variation. For the same
model, bet-hedging is never an advantageous strategy in the well-mixed
case, supporting the view that range expansions strongly promote
diversification. We further find that spatial environmental
variations provide more opportunities for bet-hedging than temporal
fluctuations.  Finally, we show that our conclusions still hold when
considering stochastic effects on the front propagation induced by a
finite population size.

The paper is organized as follows. In Section \ref{sec:model} we
introduce a general population model and an example with two available
phenotypes and two environmental states. Section \ref{sec:simple}
presents an extensive study of this example. In \ref{sec:general}, we
demonstrate that the main conclusions obtained for the example also
hold for the general model. Section \ref{sec:discussion} is devoted to
conclusions and perspectives.

\section{Model}\label{sec:model}

\begin{figure}[h!]
 \centering
 \includegraphics[width=0.9\textwidth]{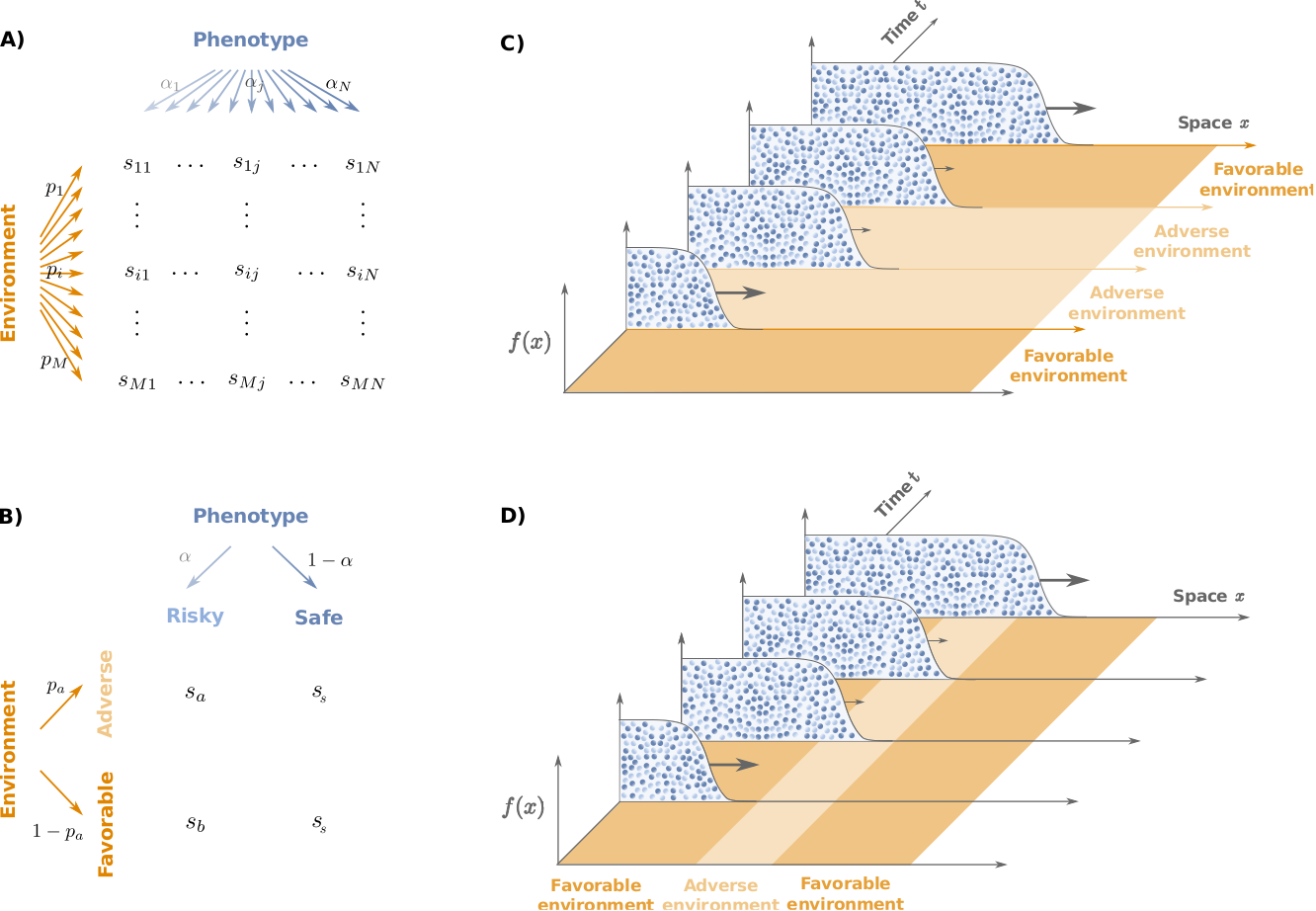}
\caption{\textbf{Population model}. A) General model: individuals can adopt 
$N$ different phenotypes with probabilities $\alpha_j$ ($j=1,\cdots,N$) and experience $M$
different environmental conditions with probabilities $p_i$ ($i=1,\cdots,M$). 
The fitness of an individual with phenotype $j$ in an environment $i$ is given by $s_{ij}$.
B) Two-phenotypes model: Individuals can adopt either a
  ``risky'' or a ``safe'' phenotype with probabilities $\alpha$, and
  $1-\alpha$ respectively. The safe phenotype is characterized by an
  environment-independent growth rate $s_s$. The growth rate of the risky phenotype is $s_a$ or
  $s_b$, depending on whether the current environment is
  ``adverse'' (a) or ``favorable'' (b).  C) and D)
  Sketch of range expansion in a population having $0\le\alpha\le1$
  for temporally varying C) and spatially varying D) environments,
  respectively.}
  \label{fig:model}
\end{figure}

We consider a population consisting of individuals that can assume $N$
alternative phenotypes. The population as a whole adopts a phenotypic
strategy, that is identified by the fractions $\alpha_i$, $i=1\dots N$
of the population assuming each phenotype $i$ with $\sum_i \alpha_i=1$
and $0\le\alpha_i\le1$ $\forall i$ (Fig. \ref{fig:model}A). As customary in game theory, we
say that a strategy is a ``pure strategy'' if $\alpha_i=\delta_{ik}$
for some phenotype $k$, and a ``mixed strategy'' otherwise.  We assume that the
$\alpha_i$'s remain constant in time within the population.

The environment can be found in one of $M$
different states, which can randomly alternate either in time or in
space. We call $p_i$ the probability of encountering environment $i$.
We further define the growth rate $s_{ij}\ge 0$ of phenotype $j$ in
environment $i$ (Fig. \ref{fig:model}A). When the population size is sufficiently large, so
that demographic stochasticity can be neglected, the
population-averaged growth rate given the state $i=i(x,t)$ of the environment
at position $x$ and time $t$ is

\begin{equation}\label{growthrate_general}
\sigma_i=\sum_j \alpha_j s_{ij}.
\end{equation}

Since Eq. \ref{growthrate_general} is linear in the $\alpha_j$'s,
the population-averaged growth rate in a given environment is always
maximized by the pure strategy with the highest growth rate. However,
in the presence of uncertainty about the environment, the population
might choose other strategies. One possibility is to select a
different pure strategy, less
risky than the optimal one. This case is often termed ``conservative
bet-hedging'' in the ecological literature
\citep{hopper1999risk}. Another option is to adopt a mixed strategy,
with different phenotypes more adapted to different environments.
This case is termed ``diversifying bet-hedging'' in the literature
\citep{den1968spreading,hopper1999risk}. Since our interest is in
diversification, the term ``bet-hedging'' will refer herein to
diversifying bet-hedging.

Before presenting our results in full generality, we will illustrate
it in a simple, yet ecologically relevant 
instance of the model with
only two phenotypes: ``safe'' and ``risky'' and two environmental
states: ``adverse'' (a) and ``favorable'' (b). The safe phenotype is
characterized by a growth rate $s_s$ both in the adverse and favorable
environments. The growth rate of the risky phenotype is $s_a$ in
environment (a) and $s_b$ in environment (b) (Fig. \ref{fig:model}B) \citep{hufton2018phenotypic}.
The two environments occur with the same probability, $p_a=p_b=1/2$. A
fraction of individuals $\alpha$ adopts the risky phenotype and the
complementary fraction $(1-\alpha)$ adopts the safe phenotype
(Fig. \ref{fig:model}B). For this model, the population-averaged
growth rate reads

\begin{equation}\label{eq:sigma}
 \sigma(x,t) = \begin{cases} \sigma_a= (1-\alpha)s_s+\alpha s_a,
 & \mbox{in env. a} \\ \sigma_b= (1-\alpha)s_s+\alpha s_b . & \mbox{in env. b} \end{cases}
\end{equation}

Note that, with a slight abuse of notation, we use equivalently $\sigma_i$
or $\sigma(x,t)$ to denote the population-averaged growth rate in the
environment $i(x,t)$. For pure strategies, $\alpha=0$ or $\alpha=1$, the population-averaged
growth rate $\sigma$ reduces to the growth rate of the safe or risky
phenotype, respectively.

\section{Results}\label{sec:modelandresults}

\subsection{Two-phenotype model}\label{sec:simple}

We seek to understand those conditions
under which bet-hedging is advantageous for the population.
To this end, we shall compare three situations: i) well-mixed
populations, ii) range expansions in environments that fluctuate
temporally, but that are homogeneous in space
(Fig. \ref{fig:model}C), and iii) range expansions in
spatially fluctuating environments that are homogeneous in time
(Fig. \ref{fig:model}D).

\subsubsection{Well-mixed case} \label{sec:kelly}

We start by analyzing the well-mixed case, where the
spatial coordinates of individuals can be ignored. The total
population density $f(t)$ evolves according to the
equation 

\begin{equation}\label{eq:mf}
\frac{d}{dt}f(t)=\sigma(t) f(t) .
\end{equation}

In writing Eq. \ref{eq:mf}, we used the assumption that the fraction
$\alpha$ of the population adopting the risky phenotype remains
constant in time (see \citep{ashcroft2014fixation,hufton2016intrinsic}
for cases in which this assumption is relaxed). Equation \ref{eq:mf}
can be readily integrated, obtaining

\begin{equation}\label{eq:mf_sol}
\ln\left(\frac{f(t)}{f(0)}\right)=\int_0^t
dt'~\sigma(t')\stackrel{t\gg1}{\longrightarrow}t\langle \sigma_i\rangle
\end{equation}

where $\langle \sigma_i\rangle=\sum_i p_i \sigma_i$ denotes an average
over the environmental states. For Eq. \ref{eq:mf_sol} to hold, we
do not need to make strong assumptions about the statistics of the
environmental states, other than it should be stationary, ergodic, and with a finite correlation time.

The optimal strategy $\alpha^*$ is obtained by maximizing the
right-hand side of Eq. \ref{eq:mf_sol} respect to the strategy
$\alpha$.  Since $\langle \sigma_i\rangle$ is a linear function of
$\alpha$, its maximum is always reached at the extremes of the
interval ($\alpha \in [0,1]$). In particular, defining the normalized
growth rates $\tilde{s}_a\equiv s_a/s_s$ and
$\tilde{s}_b \equiv s_b/s_s$, we find that the optimal strategy is
$\alpha^*=1$ when $\tilde{s}_b > 2-\tilde{s}_a$ and $\alpha^*=0$
otherwise. This means that no bet-hedging strategy is possible in this
model in the well-mixed case \citep{hufton2018phenotypic}.

This simple result illustrates an aspect of bet-hedging that is
sometimes under-appreciated. In well-mixed systems, bet-hedging
optimal strategies appear when the model includes at least one of the
following ingredients: a) discrete generations, as in the seminal work
of Kelly \citep{kelly2011new}, b) finite switching rates among
strategies \citep{kussell2005phenotypic,hufton2016intrinsic}, or c) a
delta-correlated environment \citep{hidalgo2015stochasticity}. Any of
these ingredients can lead to nonlinearities in the average
exponential growth rate, therefore opening the way for a non-trivial
optimal strategy.

Note that, in this model, the frequency of environmental change
does not play a role, as far as it is finite
\citep{hidalgo2015stochasticity}. The physical reason can be understood from
the right-hand side of Eq. \ref{eq:mf_sol}: the optimal strategy depends on the
frequency of different environmental states but not on the switching
rates. This feature is also shared by other well-mixed models that do
allow for optimal bet-hedging strategies, such as the classic model by
Kelly \citep{kelly2011new}.
We shall see in the following that, on the contrary, the rate of
environmental change plays an
important role for expanding populations.

\subsubsection{Range expansion in fluctuating
  environments} \label{sec:space}
We now consider a population expanding into an unoccupied,
one-dimensional space under the influence of a stochastically changing
environment. Its population dynamics are described by the Fisher
equation \citep{Fisher,Saarloos}:
\begin{equation}\label{eq:fisher}
\partial_t f(x,t)= D\nabla^2 f(x,t) + \sigma(x,t) f(x,t)(1-f(x,t)),
\end{equation}
where $f(x,t)$ is the population density at spatial coordinate $x$ and
time $t$, and $D$ is the diffusion constant, which characterizes the
motility of individuals.  For a constant growth rate $\sigma$, the
stationary solution of Eq. \ref{eq:fisher} is characterized by a front
advancing in space with velocity $v_F=2\sqrt{D\sigma}$.  
Instead, we consider a fluctuating case in which the growth rate $\sigma(x,t)$
depends on the population strategy $\alpha$ and on environmental
conditions according to Eq. \ref{eq:sigma}. In such case, one can define an asymptotic
mean velocity of the front as
\begin{equation}\label{eq:vm}
v_M = \lim_{t\rightarrow\infty}\frac{1}{t}\int_{0}^\infty f(x,t)~dx.
\end{equation}
In what follows, we take $v_M$ as a proxy of the long-term population
fitness and maximize it with respect to $\alpha$ to determine the
optimal strategy.

\subsubsection{Range expansion in temporally varying
  environments} \label{sec:temporal_var} We first consider the case in
which environmental conditions change randomly with time, but are
homogeneous across space, $\sigma(x,t)=\sigma(t)$ (see
Fig.\ref{fig:model}C). Switching rates between adverse and favorable
environments are $k_{a\to b}=k_{b\to a}=k$. We first estimate the
asymptotic mean velocity defined in Eq. \ref{eq:vm} in the limiting
cases of $k \to 0$ and $k \to \infty$.

\begin{figure}[h!]
\centering
 \includegraphics[width=0.8\textwidth]{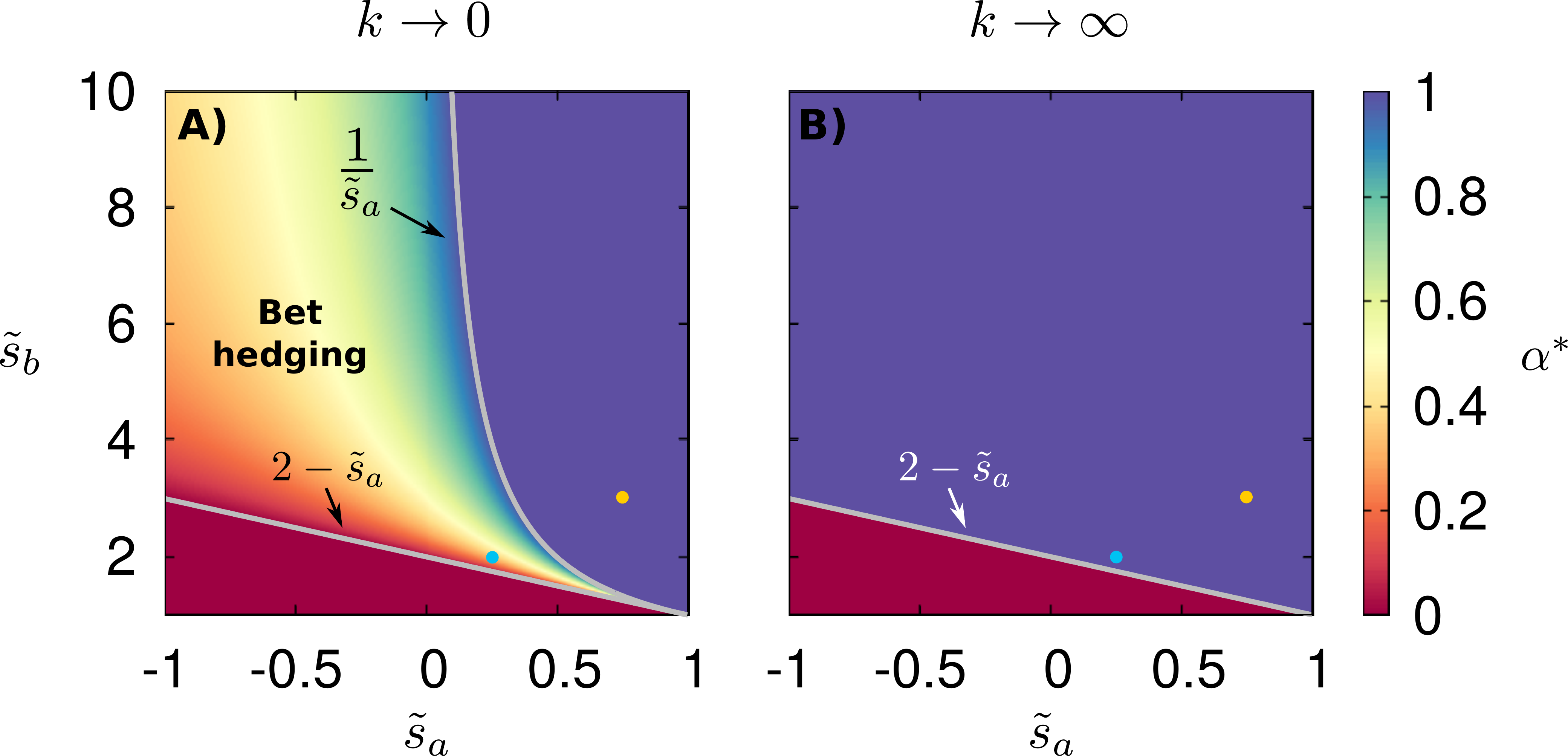}
\caption{
\textbf{Bet-hedging region in temporally varying environments}
Optimal strategy $\alpha^*$ as a function of growth rates
$\tilde{s}_a\equiv s_a/s_s$ and $\tilde{s}_b \equiv s_b/s_s$ for range expansions
in temporally varying environments under the limits of
environmental change rate (A) $k\rightarrow 0$, see
Eq.\ref{eq:fisher_conditions1},  and (B) $k\rightarrow \infty$. In all panels, lines delimit
the bet-hedging region $0\le \alpha^*\le 1$. Two dots in the panels 
mark parameter values chosen for the analysis of Figs.
\ref{fig:vel_integration_temporal},\ref{fig:vel_integration_spatial},\ref{fig:vel_stochastic_temporal}.
}
\label{fig:alphaopt}
\end{figure}

When the environment changes very infrequently, $k \to 0$, the population
front has the time to relax to the asymptotic shape characterized by
its corresponding Fisher velocity, $v_a=2\sqrt{D\sigma_a}$ or
$v_b=2\sqrt{D\sigma_b}$ depending on the environment
\citep{Fisher,Cencini}.  Thus, the asymptotic mean velocity can be
estimated as $ v_M=(v_a+v_b)/2$. Maximizing $v_M$ with respect to
$\alpha$, we find that in this case, a bet-hedging optimal strategy
exists under the conditions  (Fig. \ref{fig:alphaopt}A):
\begin{eqnarray}\label{eq:fisher_conditions1}
 \tilde{s}_b &>& 2-\tilde{s}_a,\nonumber\\
 \tilde{s}_b&<&1/\tilde{s}_a .
\end{eqnarray}
In the opposite limiting case of a rapidly fluctuating environment,
$k\to \infty$, the population effectively experiences the average of
the two growth rates, so that the velocity can be estimated
as $v_M\approx 2\sqrt{D\langle\sigma\rangle}$, where
$\langle\dots\rangle$ denotes an average over the environmental
states. In this case, the optimal strategy $\alpha^*$ is achieved by
maximizing the average growth rate $\langle\sigma\rangle$ with respect
to $\alpha$. Since $\langle\sigma\rangle$ is linear in $\alpha$, the
maximum always lies at the extremes of the interval $[0,1]$. In
particular, we find $\alpha^*=1$ when $ \tilde{s}_b > 2-\tilde{s}_a$
and $\alpha^*=0$ otherwise, as in the well-mixed case.  This implies
that no bet-hedging regime exists in this limit
(Fig. \ref{fig:alphaopt}B).

\begin{figure}[h!]
 \centering
 \includegraphics[width=0.8\textwidth]{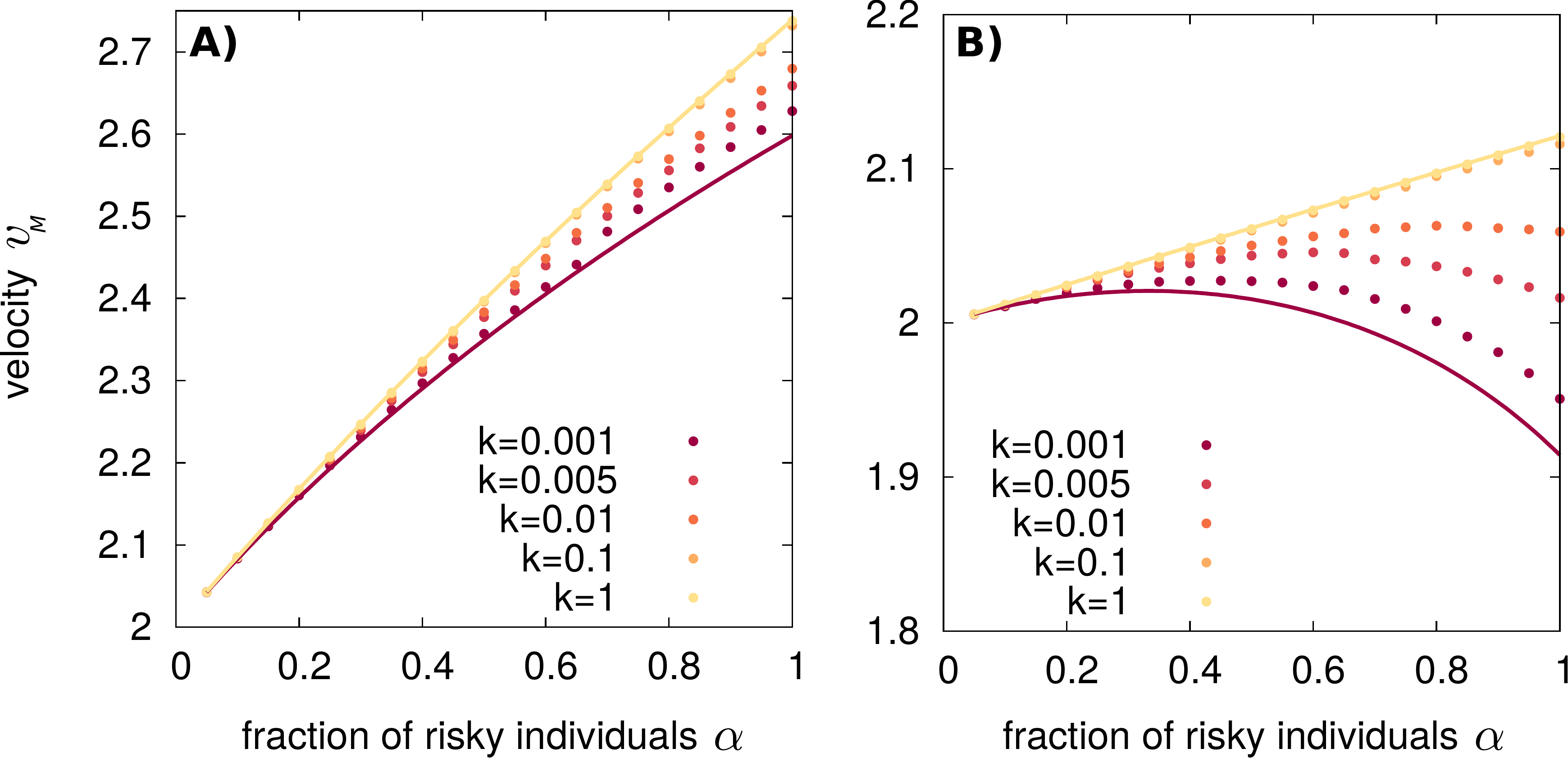}
\caption{\textbf{The asymptotic mean velocity increases with $k$ in
    temporally varying environments.}  (A) Velocities obtained by
  numerical integration of Eq. \ref{eq:fisher} for $s_a=0.75$,
  $s_s=1$, $s_b=3$ (yellow dot of Fig. \ref{fig:alphaopt}) for
  different switching rates $k$ shown in the figure legend. (B) The same
  for $s_a=0.25$, $s_s=1$, $s_b=2$ (blue dot of
  Fig. \ref{fig:alphaopt}).  In (A), the optimal
  strategy is $\alpha=1$ for all $k$ values. In (B),
  bet-hedging optimal strategies appear depending on the
  value of $k$.  The continuous red and
  yellow lines (both in A and B) illustrate analytical predictions under the two limits
  $v_M (k\rightarrow 0) = (v_a(\alpha)+v_b(\alpha))/2$  and
  $v_M(k\rightarrow \infty)=2\sqrt{D\langle\sigma(\alpha)\rangle}$,
  respectively.
  \label{fig:vel_integration_temporal}}
\end{figure}

To explore the intermediate regimes of finite $k$, it is necessary to
resort to numerical simulations of Eq. \ref{eq:fisher}.  For a set of
parameters such that the optimal strategy is $\alpha^*=1$ for
$k\rightarrow 0$, the optimal strategy remains $\alpha^*=1$ for all
values of $k$, see Fig. \ref{fig:vel_integration_temporal}A. Instead,
in a case where the optimal solution is in the bet-hedging region for
$k\rightarrow 0$, the optimal strategy $\alpha^*$ increases with the
switching rate, so that for large $k$ the optimal strategy is outside
the bet-hedging region, $\alpha^*=1$. These results support our
analytical estimates of limiting values and suggest that the
asymptotic mean velocity is a monotonically increasing function of the
switching rate $k$ in this case.

\subsubsection{Range expansion in spatially varying
  environments} \label{sec:spatial_var} We now consider the case in
which environmental conditions are constant in time, but depend on the
spatial coordinate $x$. The dynamics are described by the Fisher
equation \ref{eq:fisher} with two types of environment randomly
alternating in space, $\sigma(x,t)=\sigma(x)$. We call $k_S$ the
spatial rate of environmental switch, so that the probability of
encountering an environmental shift within an infinitesimal spatial
interval $dx$ is equal to $k_S~dx$. The switching rates from
environment $a$ to $b$ and vice-versa are both equal to $k_S$. As
above, we first analyze the two limits $k_S \to 0$ and
$k_S \to \infty$.

In the limit $k_S\rightarrow 0$, the population front traverses
large regions of space characterized by a constant environment, either
$a$ or $b$, thus being able to reach the corresponding Fisher velocity,
$v_a$ or $v_b$, respectively.  The mean traversed lengths $\Delta x_a$ and $\Delta x_b$ are equal for the
two environments.  On the other hand, the mean times spent in each of them,
$t_a$ and $t_b$, are different, and satisfy the relation

\begin{equation}
 \displaystyle\frac{t_a}{t_b}=\displaystyle\frac{\Delta
   x_a/v_a}{\Delta x_b/v_b}=\displaystyle\frac{v_b}{v_a}.
\end{equation}
Therefore, in this case, the asymptotic mean velocity is given by the
harmonic mean of the velocities in the two environments
\begin{equation} \label{eq:vel_spatial}
 v_M (k_S\to 0)=\displaystyle\frac{t_av_a+t_bv_b}{t_a+t_b}=\displaystyle\frac{2v_av_b}{v_a+v_b} .
\end{equation}

At the opposite limit of large $k_S$, the environment is characterized
by frequent spatial variations. In this case, the population front
occupies multiple $a$ and $b$ sectors with an effective growth rate
$\langle\sigma\rangle$.  As in the time-varying case, the asymptotic
mean velocity in this limit is
$v_M (k_S\to\infty)= 2\sqrt{D\langle\sigma\rangle}$, see also
\citep{shigesada1986traveling,shigesada1997biological}.

\begin{figure} [h!]
 \centering
 \includegraphics[width=0.8\textwidth]{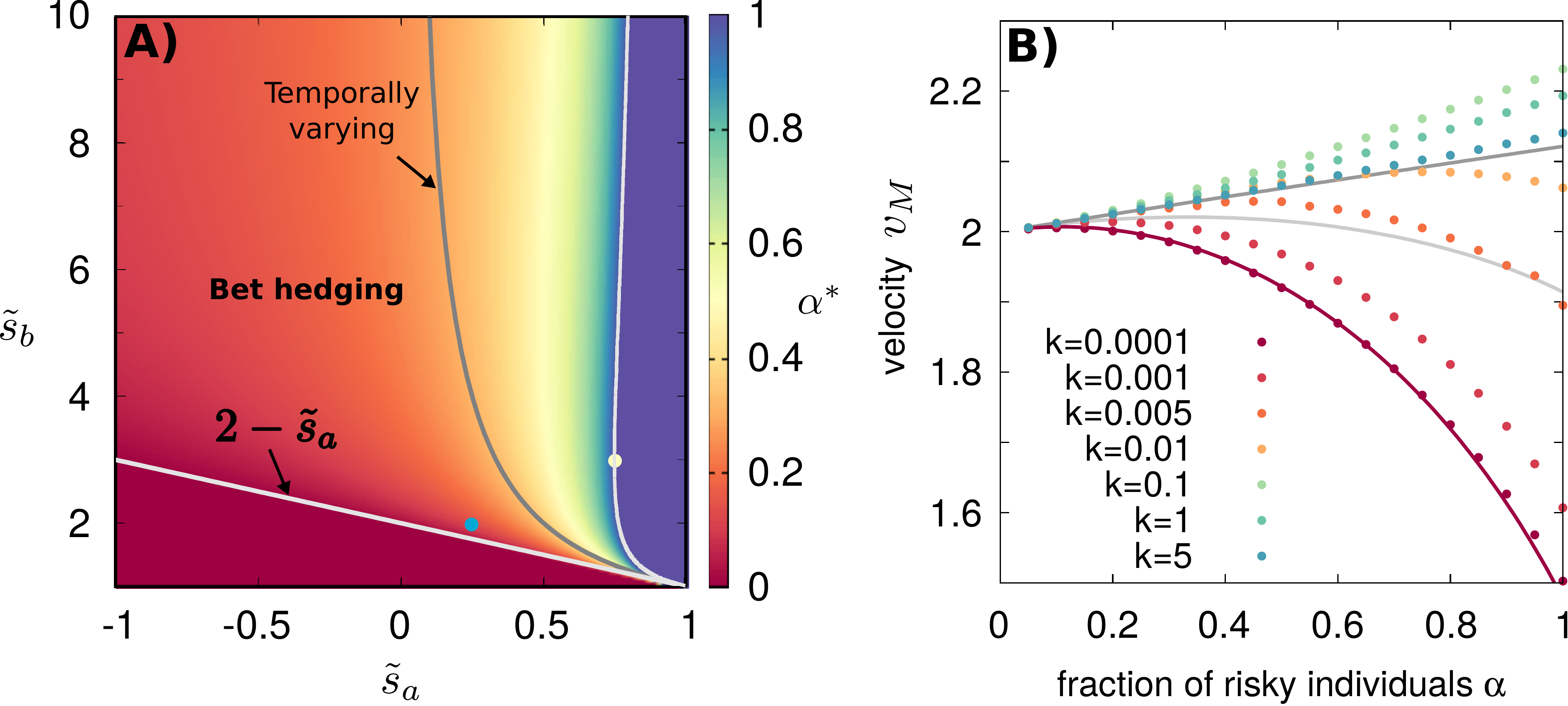}
\caption{\textbf{The bet-hedging region is expanded for range
    expansions in spatially varying environments compared to
    temporally varying environments.}  A) Optimal strategy $\alpha^*$ as a
  function of the parameters for spatially varying environments in the
  limit $k_s\to 0$, Eq. \ref{eq:vel_spatial}. White lines mark the
  limits of the bet-hedging region. The limit for which the
  strategy $\alpha=1$ is optimal in temporally fluctuating 
  environments for $k\rightarrow 0$ is also shown  (gray line)
  for comparison. B) The velocity obtained by numerical integration of
  Eq. \ref{eq:fisher} for $s_a=0.25$, $s_s=1$, $s_b=2$ (corresponding to
  the blue dot of panel A) and different values of
  $k_S$ shown in the figure legend. Light and dark gray lines
  correspond to the analytical limits for temporally varying
  environments, $v_M (k\rightarrow 0)=(v_a(\alpha)+v_b(\alpha))/2$,
  and
  $v_M(k\rightarrow \infty)=v_M(k_S\rightarrow
  \infty)=2\sqrt{D\langle\sigma(\alpha)\rangle}$, respectively. The
  red curve is the analytical solution for a spatially fluctuating
  environment with $k_S\to 0$, see Eq. \ref{eq:vel_spatial}. Note that
  in this case, the asymptotic mean velocity does not increase
  monotonically with $k_S$ but is maximal at $k_S\approx 0.1$.}
\label{fig:vel_integration_spatial}
\end{figure}

Here, for $k_S\rightarrow 0$ the bet-hedging region is broader with
respect to the temporally fluctuating environment for
$k\rightarrow 0$, see Fig. \ref{fig:vel_integration_spatial}A. For
$k_S\to\infty$, the optimal strategy is the same as in
Fig. \ref{fig:alphaopt}C and there is no bet-hedging regime.

We numerically solved Eq. \ref{eq:fisher} for intermediate values of
$k_S$ and obtained the mean asymptotic velocities as a function of
$\alpha$, see Fig.  \ref{fig:vel_integration_spatial} B. Results
support theoretical predictions in the limiting cases
$k_S\rightarrow 0$ and $k_S\rightarrow \infty$. In this case, 
we observe a non-monotonic behavior of $v_M$ as a function of $k_S$, 
so that the maximum mean velocity is attained at an
intermediate switching rate. An analytical explanation of this
non-trivial effect goes beyond the scope of this work.

 \subsubsection{Effect of finite population size} \label{sec:stochastic} 
 The deterministic Fisher equation (\ref{eq:fisher}) is rigorously
 valid only in the limit of infinite local population sizes. We now
 explore the robustness of our results when considering stochasticity
 induced by the finite size of populations, i.e. ``demographic
 noise''.  We focus on the case of a front propagating in a
 temporally varying environment. To study finite population size, we
 solve numerically a stochastic counterpart of the Fisher equation
 \begin{equation}\label{eq:fisher_stoch}
\dot{f}(x,t)= D\nabla^2 f + \sigma(t)
f(1-f)+\sqrt{\frac{2}{N}f(1-f)}\xi(x,t) ,
\end{equation}
see e.g. \citep{korolev2010genetic}. In Eq. \ref{eq:fisher_stoch},
$\xi(x,t)$ is Gaussian white noise with $\langle \xi(x,t)\rangle=0$,
$\langle \xi(x,t) \xi(x',t')\rangle=\delta(x-x')\delta(t-t')$.  The
parameter $N$ represents the number of individuals per unit length
corresponding to $f(x,t)=1$. For large population sizes, $N\gg 1$,
Eq. \ref{eq:fisher_stoch} reduces to Eq. \ref{eq:fisher}.  Numerical
integration of Eq. \ref{eq:fisher_stoch} requires some care due to the
fact that both noise and the deterministic terms go to zero as the
absorbing states $f(x,t)=0$ and $f(x,t)=1$ are approached
\citep{dornic2005integration,Moro2}. A detailed description of our
integration scheme is presented in the Supporting information.

\begin{figure} [h!]
\centering
 \includegraphics[width=0.8\textwidth]{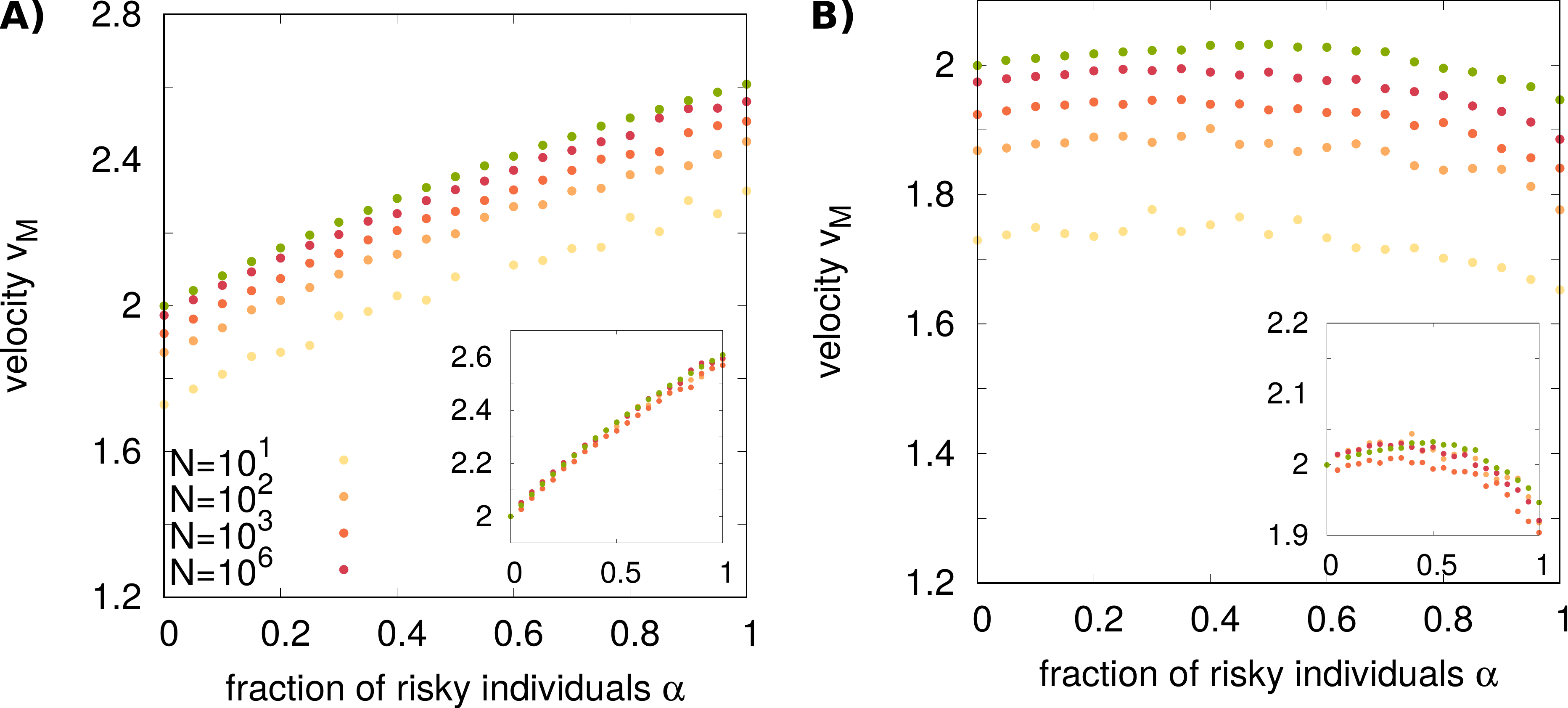}
\caption{ \textbf{The optimal strategy is robust with respect to
    noise induced by finite population size in
    temporally varying environments.}  (A) Asymptotic mean velocities
  obtained by numerical integration of the stochastic Fisher
  equation \ref{eq:fisher_stoch} for $\tilde{s}_a=0.75$, $s_s=0.01$,
  $\tilde{s}_b=3$ (yellow dot of Fig. \ref{fig:alphaopt}) and
  different population sizes. (B) The same for   $\tilde{s}_a=0.25$, $s_s=1$, $\tilde{s}_b=2$ (blue dot of
  Fig. \ref{fig:alphaopt}).  In both panels, the temporal switching
  rate of the environment is $k=0.001$.  Green dots corresponds to the
  results of Figs. \ref{fig:vel_integration_temporal}A,B for
  $k=0.001$. Insets show a collapse of the curves according to Eq. 
  \ref{scaling_derrida}, with a fitted value of $C=3$.}
\label{fig:vel_stochastic_temporal}
\end{figure}

 For a Fisher wave propagating in a homogeneous environment,
 demographic noise leads to a reduced front velocity $v$ with respect
 to the deterministic case \citep{brunet2001effect,Saarloos,Moro,Moro2}
\begin{equation}\label{scaling_derrida}
(v-v_F)\sim -\frac{C}{\ln^2(N)}
\end{equation}
where $C$ is a constant, $N$ is the maximum population size 
per unit length, and $v_F=2\sqrt{D\sigma}$ is the Fisher velocity in the
absence of demographic noise.
Asymptotic mean velocities for stochastic waves in temporally varying
environments are shown in Fig. \ref{fig:vel_stochastic_temporal}. Also
in this case, small populations, subject to relatively strong
demographic noise, propagate more slowly than large populations. In particular,
 curves at different values of $N$ can be approximately rescaled using
Eq. \ref{scaling_derrida}, assuming that $C$ does not depend on
$\alpha$ (insets of Fig. \ref{fig:vel_stochastic_temporal}). These results imply that the optimal strategy $\alpha^*$ is
robust with respect to demographic noise, at least for moderately to
relatively large values of $N$. The same scaling holds for
spatially varying environments, but with mild deviations that seem to
expand the bet-hedging region even further, compared with the
infinite population size limit (see Supporting information).

\subsection{General bet-hedging model}\label{sec:general}

In this section, we demonstrate that our main conclusions hold also
for the general case with $N$ phenotypes and $M$ environmental states
(see Section \ref{sec:model}).  In particular, for a temporally
fluctuating environment in the limit of very slow switching rates, the
bet-hedging regime occupies a reduced region of parameter space
compared to temporally constant
environments fluctuating slowly in space.  Also in this case, we find
that for frequent environmental change, the propagation velocity tends
to $v_M\approx2\sqrt{D\langle \sigma\rangle}$, regardless of whether the
environmental fluctuations depend on time or space. Therefore, the
optimal strategy maximizes the linear function of the $\alpha_i$s 
$\langle \sigma\rangle$ and is therefore a pure strategy as
discussed after Eq. \ref{growthrate_general}.

We consider a range expansion where the environment fluctuates
in time and the stochastic switching rates among the $M$ environmental
states are small. Following the same line of thought of Section \ref{sec:temporal_var}, the optimal strategy maximizes

\begin{equation}\label{max:f}
\sigma_T=\frac{v_M(k\rightarrow 0)}{2\sqrt{D}}=   \sum_i p_i \sqrt{\sigma_i  }
\end{equation}

where, as usual, $\sigma_i=\sum_j s_{ij}\alpha_j$.
For spatially varying environments, the optimal strategy maximizes the
harmonic mean

\begin{equation}\label{max:fs}
\sigma_S=\frac{v_F(k_S\rightarrow 0)}{2\sqrt{D}} =\frac{1}{\sum_i p_i \displaystyle\frac{1}{\sqrt{\sigma_i}} }.
\end{equation}

Both for Eq. \ref{max:f} and Eq. \ref{max:fs}, maximization has to
be performed with the constraint $\sum_j \alpha_j=1$ and
$0\le \alpha_j\le 1$ $\forall j$. We recall that the
bet-hedging regime is the region of parameter space where the optimal
solution is a mixture of all phenotypes, $\alpha_i>0$ $\forall
i$. Here we show that if, for a given choice of the $s_{ij}$'s
and $p_i$'s, a population advancing in a temporally varying
environment is in a bet-hedging regime, then the same holds for 
spatially varying environments. For the demonstration, we borrow a mathematical
tool from evolutionary game theory \citep{hofbauer1998evolutionary}. We
introduce the gradients

\begin{eqnarray}
F^{T}_l&=&\frac{\partial \sigma^{T}}{\partial\alpha_l}=\left\langle\frac{
            s_l}{2\sqrt{\sigma}}\right\rangle\nonumber\\
F^S_l&=&\frac{\partial \sigma^S}{\partial\alpha_l}=(\sigma^S)^2
\left\langle\frac{s_l}{2\sigma^{3/2}} \right\rangle
\end{eqnarray}
where  $\langle x\rangle=\sum_i p_i x_i$ is the average over
environments. We now associate replicator equations to Eq. \ref{max:f} and Eq. \ref{max:fs}:

\begin{eqnarray}\label{eq:replicator}
\frac{d}{dt}\alpha_l&=&\alpha_l (F_l^T-\bar{F}^T)=\alpha_l \left\langle\frac{
            s_l-\sigma}{2\sqrt{\sigma}}\right\rangle\\
\frac{d}{dt}\alpha_l&=&\alpha_l(F^S_l -\bar{F}^S_l)=\alpha_l(\sigma^S)^2  
\left\langle\frac{s_l-\sigma}{2\sigma^{3/2}} \right\rangle.
\end{eqnarray}
The system is in a bet-hedging regime when the replicator equations
 admit a stable fixed point in the interior of
the unit simplex, $0<\alpha_i<1$.  Instead of computing the fixed
point explicitly, we check whether each phenotype $l$ has a
positive growth rate for $\alpha_l\ll 1$. Brouwer's fixed
point theorem ensures that, under this condition, there must be a
fixed point in the interior, see
\citep{hofbauer1998evolutionary}, chapter 13.
For our aims, it is therefore sufficient to prove that, for small
$\alpha_l$, if $(F_l^{T}-\bar{F}^{T})$ is positive, then
$ (F_l^{S}-\bar{F}^{S})$ must be positive as well. Note that
for $\alpha_l\ll 1$, the average
$\sigma=\sum_j s_{ij} \alpha_j$ does not depend on $\alpha_l$,
and therefore, $\sigma$ and $s_l$ are uncorrelated random
variables respect to the average over the environment.
Since $\sqrt{\sigma}>0$, this means that the sign of $(F_l^T-\bar{F}^T)$ is the same than the quantity
\begin{equation}\label{step}
\frac{1}{\langle\sqrt{\sigma}\rangle}\left\langle s_l\right\rangle \left\langle\frac{1}{\sqrt{\sigma}}\right\rangle-1 .
\end{equation}
Following the same logic, the sign of $(F_l^S-\bar{F}^S)$ is the same than

\begin{equation}
  \left\langle s_l \right\rangle
  \left\langle\frac{1}{\sigma^{3/2}}\right\rangle-\left\langle\frac{1}{\sqrt{\sigma}}
  \right\rangle= \left\langle\frac{1}{\sqrt{\sigma}}\right\rangle
  \left( \displaystyle\frac{\left\langle s_l \right\rangle\left\langle
        1/\sigma^{3/2}\right\rangle}{\left\langle
        1/\sqrt{\sigma}\right\rangle}-1 \right) .
\end{equation}

Since also $\langle s_l\rangle>0$, we need to demonstrate that the following inequality always holds

\begin{equation}
\displaystyle\frac{\left\langle 1/\sigma^{3/2}\right\rangle}{\left\langle 1/\sqrt{\sigma}\right\rangle} \ge 
\left\langle\frac{1}{\sqrt{\sigma}}\right\rangle\frac{1}{\langle\sqrt{\sigma}\rangle}.
\end{equation}

This can be proven from the chain of inequalities

\begin{equation}\label{eq:chain}
\displaystyle\frac{\left\langle 1/\sigma^{3/2}\right\rangle}{\left\langle 1/\sqrt{\sigma}\right\rangle}\geq\left\langle \frac{1}{\sigma}\right \rangle\ge \left\langle
    \frac{1}{\sqrt{\sigma}}\right \rangle
\left\langle
    \frac{1}{\sqrt{\sigma}}\right \rangle
\ge
\left\langle  \frac{1}{\sqrt{\sigma}}\right \rangle
    \frac{1}{\left\langle\sqrt{\sigma}\right\rangle} .
\end{equation}

In Eq. \ref{eq:chain}, the second and third inequalities are
consequences of Jensen's inequality, since both $x^2$ and $1/x$ are
convex functions. For the first inequality in Eq. \ref{eq:chain},
since $s>0$, we can  use the result $\left\langle x^{i} \right\rangle  \geq \left\langle x^{j} \right\rangle^{i/j}$ proved for $i>j$ in \citep{kapur1995testing}.
Combining this result for $(i=3,j=2)$ and $(i=2,j=1)$, we obtain $ \left\langle x^3 \right\rangle \geq \left\langle x^2 \right\rangle \left\langle x \right\rangle$.
Taking $\left\langle x \right\rangle=\left\langle 1/\sqrt{\sigma}
\right\rangle$ we finally prove  Eq. \ref{eq:chain}.
Therefore, in the limit of small switching rates of the environment, 
the bet-hedging region is wider in the spatially varying case than 
in the temporally varying case.

In the opposite limit of high rates of environmental switch, the function
to be optimized is linear, and the optimal strategy is a pure
strategy. In this case, the particular phenotype $l$ adopted by the whole
population is that maximizing $\sum_i p_i
s_{il}$. This conclusion holds both for temporally and spatially
varying environments.

\newpage
\section{Conclusions}\label{sec:discussion}

Understanding the precise mechanisms of population expansions is of
utmost importance, not only for understanding species diversity, but
also to cope with invasive species in new habitats
\citep{wolf2012animal,sih2012ecological,chapple2012can,carere2013animal},
bacterial infections
\citep{frankel2014adaptability,dufour2014limits,dufour2016direct,jones2010dormancy},
and cell migration, such as those occurring during tissue renewal or
cancer metastasis \citep{mayor2016front}.  Phenotypic diversity is a
convenient strategy for the success of population expansions in a
broad range of contexts \citep{wolf2012animal,
  sih2012ecological,chapple2012can,carere2013animal,
  frankel2014adaptability,dufour2014limits,dufour2016direct}.
Although precise experimental measures are not easy to obtain, a
recent study shows that populations with increased variability in
individual risk-taking can colonize wider ranges of territories
\citep{moller2012between}.

In this work, we proposed a general mathematical and computational
framework to analyze such scenarios. In particular, we introduced a
population model with diverse phenotypes that perform differently
depending on the type of environment. We focused on the ``optimal''
degree of diversity leading to the fastest average population
expansion in an environment fluctuating either in space or in time. We
found that, contrarily to the well-mixed case, bet-hedging can be
convenient in expanding populations. This result complements the study
in \citep{hidalgo2015stochasticity} for a fixed habitat and supports
the view that diversification is of broad importance for
spatially-structured populations.  For environments varying slowly in
time, the expansion is relatively slow, and diverse communities can be
optimal depending on the parameters.  On the contrary, for fast
environmental changes, the optimal population always adopts a unique
strategy.

A remarkable outcome of our analysis is that spatial fluctuations
create more opportunities for bet-hedging than temporal fluctuations,
in that the region of parameter space where the optimal
population is diverse, is always larger in the former case.  One
intuitive explanation is that in the case of spatial fluctuations,
the population spends less time traversing favorable patches
than adverse ones. This means that the beneficial effect of favorable
patches is reduced with respect to the case of temporal
fluctuations. Therefore, a pure risky strategy is less efficient in
the case of spatial variability and can be more easily outcompeted by
a diversified bet-hedging strategy.

The framework presented here can be extended to accommodate other
scenarios. We have assumed that the fraction of individuals adopting
each phenotype is fixed by the phenotypic switching rates.
To understand the evolution of bet-hedging, it could be interesting to
study scenarios in which the phenotypic switching rates are slower, so
that phenotypes can be selected, and/or are themselves subject to
evolution and selection \citep{xue2016evolutionary,hufton2018phenotypic}.
Another potentially relevant extension would be to consider two-dimensional
habitats. Although the classic theory for Fisher waves
\citep{Fisher,Kolmogorov} is unaffected in higher dimensions, in the
presence of spatial heterogeneity the front shape can become
anisotropic, potentially affecting the results. 
Similarly, it would be interesting to analyze the combined effect of spatial and temporal
variability.  We also limited ourselves to the case where the
different environments affect individual growth rates, whereas in
general, one could also expect them to have an effect on motility
\citep{shigesada1979spatial,shigesada1986traveling,pigolotti2014selective,pigolotti2016competition,gueudre2014explore},
opening the way for different forms of bet-hedging.  Finally, the
present study was limited to pulled waves.  It would be interesting to
study the effect of bet-hedging on pushed waves, for example to
describe population expansion in the presence of an Allee effect
\citep{gandhi2016range,birzu2018fluctuations}.

It would be also interesting to experimentally test our
results. Experiments of expanding bacterial colonies in
non-homogeneous environments have already been performed and shed
light, for example, on the evolution of antibiotic resistance in
spatially-structured populations \citep{baym2016spatiotemporal}. To
perform experiments within the limits of our theory, a challenge can
be to maintain the environmental variability sufficiently low to avoid
exposing the population to an excessive evolutionary pressure. Similar
problems appear, for example, in studies of range expansion of
mutualistic bacteria \citep{muller2014genetic}.  An extension of the
theory including both phenotypic and genetic diversity could account
for these scenarios.

In summary, we have introduced a model to
understand conditions favoring diversification of an expanding
population. Our work provides a bridge between the theory of
bet-hedging and that of ecological range expansion described by
reaction-diffusion equations. The results of the model highlight the
relation between population diversity and fluctuations of the
environment encountered during range expansion.  The flexibility and
generality of our framework make it a useful starting point for
applications to a wide range of ecological scenarios.

\section{Acknowledgments}
We acknowledge Steven D. Aird, R. Rubio de Casas, and Massimo Cencini
for comments on a preliminary version of this manuscript. MAM is
grateful to the Spanish-MINECO/AEIa for financial support (under grant
ref.  FIS2017-84256-P; FEDER funds).

\newpage
\section{Supporting information}

\subsection{Numerical integration of the stochastic Fisher equation}

In this section we describe in detail the methods applied for the
integration of the wave equations of the two-phenotype model
studied in the Main Text.

\subsubsection{Fisher wave}

We consider the Fisher equation 

 \begin{equation}\label{eq:fisher_xt}
\dot{f}(x,t)= D\nabla^2 f(x,t) + \sigma(x,t) f(x,t)(1-f(x,t)),
\end{equation}

where $f(x,t)$ is the population density at space $x$ and time $t$, and
$\sigma(x,t)$ is the local growth rate. \\

We employ a finite-difference fourth-order Runge-Kutta method.
The systems is initialized by fixing $f(x_i,0)=1$ for $i\in(0,50)$ 
and $f(x,t)=0$ for $i>{50}$. To fix $dx$, we implement 
an adaptive routine. We intialize the routine with an 
initial guess for $dx=0.14$. Then

\begin{enumerate}
 \item We let the system evolve until the front reaches a stationary state.
 \item We compute the smallest values of $x$ for which $f(x,t)>\theta$
   for $\theta=3/4$ and $\theta=1/4$. We denote these two values as $x_{3/4}$ and $x_{1/4}$ respectively.
 \item We measure the precisions $\Delta f_{3/4}=f(x_{3/4}-dx)-f(x_{3/4})$, $\Delta f_{1/4}=f(x_{1/4}-dx)-f(x_{1/4})$.
 \item If $\Delta f_{3/4}>0.01$ and $\Delta f_{1/4}>0.01$, then $dx$ is a valid increment.
\item Otherwise, the system is set to the initial conditions and the routine is again run for $dx = d\tilde{x}-0.01$; 
being $d\tilde{x}$ the previously employed increment.
\end{enumerate}

Once $dx$ is selected, $dt$ is fixed following the Courant-–Friedrichs-–Lewy condition for an explicit integration method \citep{courant1967partial}:

\begin{equation}
 \frac{v_\mathrm{max}dt}{dx}\leq 1
\end{equation}

being $v_\mathrm{max}$ the estimated maximum velocity of the wave. We fix
$v_\mathrm{max}=100$, which is an overestimation of the maximum velocity in
our simulations.

Temporal environmental switch is numerically implemented with a simple
first-order algorithm. At the beginning of each time step, the state
of environment is switched with probability $k ~dt$. We verified that
this quantity is always sufficiently small, so that the first-order
algorithm yields reliable results. A similar algorithm is implemented for spatial environmental
variations to sequentially assign an environmental state to each
lattice site.

\subsubsection{Stochastic Fisher wave}

We consider the stochastic Fisher equation \citep{korolev2010genetic}

\begin{equation}\label{eq:fisher_stoch2}
\dot{f}(x,t)= D\nabla^2 f + \sigma(t)
f(1-f)+\sqrt{\frac{2}{N}f(1-f)}\xi(x,t) 
\end{equation}

where $\xi(x,t)$ a Gaussian white noise satisfying $\langle
\xi(x,t)\rangle=0$ and  $\langle
\xi(x,t) \xi(x',t')\rangle=\delta(x-x')\delta(t-t')$.\\

Numerical integration in the presence of noise is subtle. In
particular, one has to figure out how to deal with the unphysical
values $f(x,t)<0$ and $f(x,t)>1$ obtained numerically. In some
parameter range, the naive replacement $f(x,t)=0$ or $f(x,t)=1$ when
$f(x,t)<0$ and $f(x,t)>1$, respectively, introduces a bias that can
profoundly alter the results.  In particular, an incorrect integration
of $f(x,t)$ at the front, where
$f(x,t)$ is small, might lead to an large error in the estimated
velocity. However, when
 $f(x,t)$ is small so that
$\gamma(t)\simeq \sqrt{\frac{2}{N}f(t)}$, this problem can be
circumvented by integrating the noise term exactly
 \citep{dornic2005integration}.

\begin{figure}[htb]
\centering
\includegraphics[width=0.8\textwidth]{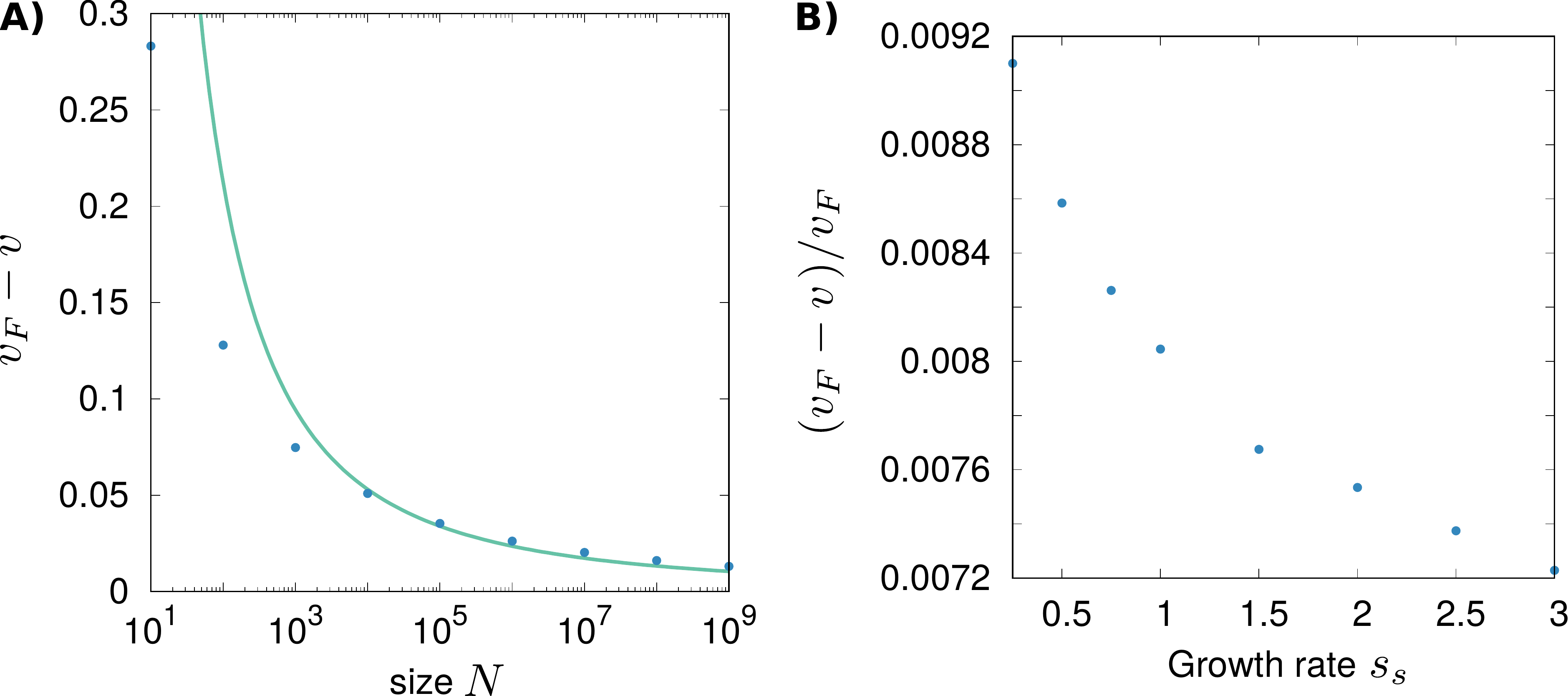}
\caption{\textbf{Size scaling and maximum error estimation for our integration method.}
Panel A) shows the curve $4.5logN^{-2}$ and the difference $v_F-v$ for $k=0$, $\alpha=0$, 
$s_s=1$, and different system sizes. Panel B) shows $(v_F-v)/v_F$ for $N=10^9$ and different growth rates
employed in this work. These results suggest that our integration method is precise and have a
maximum error of $0.0018\%$.}
\label{fig:dornic_precision}
\end{figure}

Taking this into account, we integrate the equation mixing two different
algorithms, depending on the local value of $f(x,t)$: 
\begin{itemize}

 \item If $f(x,t)>\theta$: we employ the Milstein method (order 1).  Defining
$\beta(t) \equiv D\nabla^2 f(t)+\sigma(t)f(t)(1-f(t))$ and
$\gamma(t) \equiv \sqrt{\frac{2}{N}f(t)(1-f(t))}$, the local field is
updated according to the rule

\begin{equation}
 f(x,t+dt)=f(x,t)+\beta(t) dt+\gamma(t) \Delta+\displaystyle\frac{1}{2}\gamma(t)\frac{\partial \gamma}{\partial f(t)}(\Delta^2-dt)
\end{equation}

being $\Delta=\sqrt{dt}N(0,1)$. \\

\item If $f(x,t)<\theta$: we perform the two-step numerical integration proposed in \citep{dornic2005integration}:

\begin{enumerate}
 \item {\it{Non-linear and diffusion terms}}.
 Integration of $\dot{f}(x,t)= D\nabla^2 f - \sigma(t) f^2$ is done
 by employing the Runge-Kutta method obtaining a first solution $f^*$.
 
 \item {\it{Linear and stochastic terms}}. The term
 $\sigma(t) f+\sqrt{\frac{2}{N}f}\xi(x,t)$ is integrated
 in an exact way as \citep{dornic2005integration}:
 
 \begin{equation}
  f(x,t)=r_\mathrm{Gamma}\{r_\mathrm{Poisson}\{\lambda f^*(x,t) e^{\sigma(t)t}\}\}/\lambda.
 \end{equation}

 being $\lambda=\frac{2\sigma(t)}{\gamma^2e^{\sigma(t)t}}$, and
 $r_\mathrm{Gamma}$, $r_\mathrm{Poisson}$ random values 
 obtained from the Gamma and Poisson probability distributions, i.e. 
 $Prob[r_\mathrm{Gamma}(a)=z]=\frac{z^{a-1}e^{-z}}{\Gamma [a]}$
 and $Prob[r_\mathrm{Poisson}(a)=z]=\frac{a^ze^{-a}}{z!}$, respectively. 
\end{enumerate}
\end{itemize}

To check the precision of our method we integrated the stochastic equation 
\ref{eq:fisher_stoch2} for $k=0$, $\alpha=0$, and different growth rates 
$s_s$ and compared the results to the analytical Fisher velocity $v_F=2\sqrt{Ds_s}$.
For large population size $N$, the velocity $v$ of the wave 
asymptotically goes as $v_F-v\simeq C \ln^{-2}(N)$
\citep{brunet2001effect}. Our numerical integration is consistent with
this asymptotic relation from $N\simeq 10^4$ (figure
\ref{fig:dornic_precision}A) with a root-mean-square deviation of
$0.002$. We have also obtained the values $(v_F-v)/v_F$ for the
different growth rates employed in this work to obtain an estimation
of the maximum error we expect (see figure
\ref{fig:dornic_precision}B). Note that $v_F$ is not the 
actual velocity the finite system is expected to reach,
so the relative error $(v_F-v)/v_F$ is, in fact, smaller. 
The maximum error value is around
$0.9\%$, that, considering the results of figure
\ref{fig:dornic_precision}A) leads to an
overestimated error of about $0.0018\%$.

\newpage
\subsection{Effect of finite population size for spatially varying environments}

\begin{figure}[htb]
\centering
\includegraphics[width=0.8\textwidth]{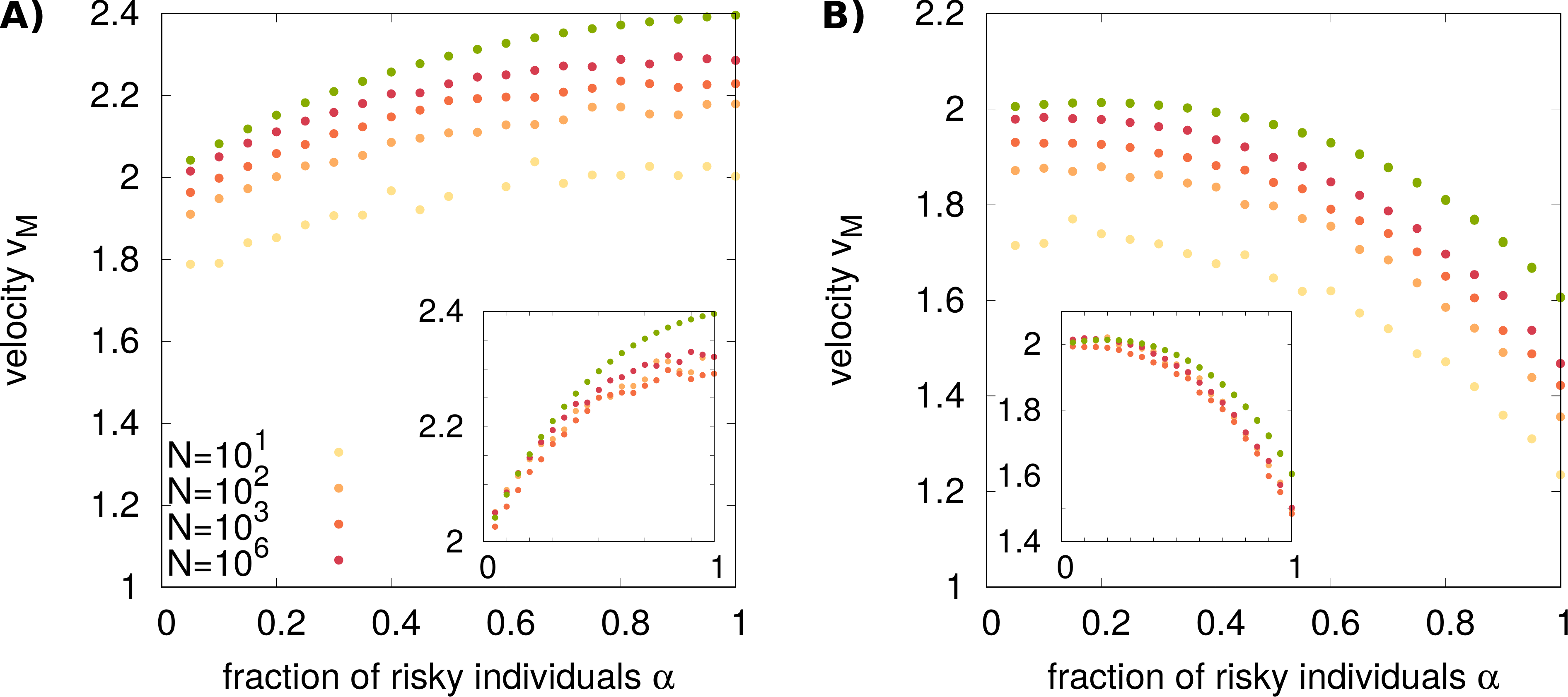}
\caption{
\textbf{Optimal strategy with
  fluctuations induced by finite population size in 
  spatially varying environments.}
  (A) Asymptotic mean velocities
  obtained by numerical integration of the stochastic Fisher
  \eqref{eq:fisher_stoch_spatial} for $\tilde{s}_a=0.75$, $s_s=0.01$,
  $\tilde{s}_b=3$ (yellow dot of Fig. 4) and
  different population sizes, shown in the figure legend. (B) Same for
  $\tilde{s}_a=0.25$, $s_s=1$, $\tilde{s}_b=2$ (blue dot of
  Fig. 4).  In both panels, the temporal switching
  rate of the environment is
  $k=0.001$.  Green dots corresponds to the results of
  the deterministic approach (Eq. 5) for $k=0.001$. Insets
  show a collapse of the curves according to Eq. 11,
  with a fitted value of $C=3$.}
\label{fig:dornic_spatial}
\end{figure}

Analogously to the temporally varying case (Section 3.1.5), we study the effect of demographic 
stochasticity induced by the finite size of the population for
spatially varying 
environments. In this case, the corresponding stochastic Fisher wave 
is described by the equation
 \begin{equation}\label{eq:fisher_stoch_spatial}
\dot{f}(x,t)= D\nabla^2 f + \sigma(x)
f(1-f)+\sqrt{\frac{2}{N}f(1-f)}\xi(x,t) ,
\end{equation}
analogous to Eq. 10 of Section 3.1.5. Following the procedure described in 
this Supplementary Material, we numerically integrated this equation. In this case,
stochasticity slightly alters the deterministic prediction (see Fig. 
\ref{fig:dornic_spatial}). Specifically, the asymptotic mean velocities $v_M$ decay 
slightly faster with the fraction of risky individuals $\alpha$. This implies
that parameters at which an optimal $\alpha^*=1$ is reached at the deterministic
approach, lead to a bet-hedging strategy with $\alpha^*<1$ in the stochastic system.
As the lower limit for stochastic systems must still be the deterministic one,
$\tilde{s}_b > 2-\tilde{s}_a$, bet-hedging region is then slightly 
enlarged for finite size populations in spatially varying environments.

Despite this slight difference between the deterministic and stochastic systems, the main
predictions described for spatially varying environments in the main text are still
maintained.

\bibliographystyle{plainnat}

\end{document}